\documentstyle[prl,aps,epsfig]{revtex}
\begin{document}

\draft
\flushbottom
\twocolumn[\hsize\textwidth\columnwidth\hsize
\csname@twocolumnfalse\endcsname

\title {
Inhomogeneous Ferromagnetism and Unconventional Charge Dynamics \\
in  Disordered Double Exchange Magnets}

\author{Sanjeev Kumar and Pinaki Majumdar }

\address{ Harish-Chandra  Research Institute,\\
 Chhatnag Road, Jhusi, Allahabad 211 019, India }

\date{May 15,  2003}

\maketitle
\tightenlines
\widetext
\advance\leftskip by 57pt
\advance\rightskip by 57pt
\begin{abstract}

We solve the double exchange  model in the presence of arbitrary substitutional
disorder by using a self consistently generated effective Hamiltonian
for the spin degrees of freedom. The magnetic properties are studied through 
classical Monte Carlo while the effective exchange, $D_{ij}$, are calculated 
by solving the disordered fermion problem, and renormalised
self-consistently with increasing temperature. We  present exact results on the 
conductivity, magnetoresistance, optical response and `real space' structure
of the inhomogeneous ferromagnetic state, and compare our results with charge 
dynamics in disordered La$_{1-x}$Sr$_x$MnO$_3$. The large sizes, ${\cal O}
(10^3)$, accessible within our method allows a complete, controlled calculation 
on the  disordered strongly interacting problem.

\

\

\end{abstract}

]

\narrowtext
\tightenlines

The double exchange (DE) model 
%\cite{old-dex}
 provides the simplest example of strong 
coupling itinerant electron ferromagnetism, and is
at the heart of the remarkable magnetotransport \cite{mang-ref}
 in the
manganites.
The qualitative magnetic features of the model are reasonably well 
understood now \cite{dex-clean-ref}, 
although transport properties have been much less explored. 
The problem is more difficult 
in the presence of quenched 
disorder, which arises inevitably in real materials  
\cite{coey,attfield,sawaki,saitoh,takenaka}
from ionic substitution and resulting bond distortions. 
Disorder leads to an 
inhomogeneous magnetic state and 
enhances the density of low energy spin  fluctuations.
The resistivity arising out of  
structural disorder and magnetic scattering is 
usually large, comparable to the Mott resistivity
\cite{sawaki,saitoh},
violates Mathiessens rule, 
and cannot be accessed by standard transport theory.
The optical response reveals a
strongly non Drude character \cite{saitoh,takenaka}, 
implying
unconventional charge dynamics, and rapid loss in low 
energy spectral weight with rising  temperature.

Most of these features, which depend explicitly on the {\it 
inhomogeneous spatial character of the magnetic and electronic
state} cannot be captured within `mean field' approximations, 
including `dynamical mean field theory' (DMFT), and 
current `real space' approaches
are severely size limited  in three dimension.
In this paper we 
use a new  \cite{sk-pm-mc}
 Monte Carlo (MC) technique, which handles 
the interplay of disorder and spin correlation essentially exactly,
to provide the first controlled results on charge dynamics 
in the disordered
double exchange (DDE) model. We map out the phase diagram, clarify the nature
of the inhomogeneous ferromagnetic state, and provide 
results on magnetotransport and optical response in the model.
Our results are directly relevant 
to the `coherent to incoherent' crossover 
\cite{sawaki,saitoh,takenaka} in disordered
R$_{1-x}$Sr$_x$Mn$_{1-z}$Al$_z$O$_3$,
and the properties of disordered metallic  
ferromagnets \cite{amorph-fm}
in general.
This approach 
can be readily extended 
to include phonon degrees of
freedom, of key importance in manganite physics.

The DDE  model has been proposed earlier to
explain the 
ferro-metal (FM) to para-insulator (PI)
transition, and the associated colossal magnetoresistance (CMR)
in manganites. 
It is now accepted that electron-phonon coupling, and possibly
phase coexistence, are important \cite{mang-ref} in understanding CMR.
While the DDE model can exhibit a FM-PI 
transition,
at large disorder, 
in real materials even weak electron-phonon
coupling will strongly affect the `Anderson localised' 
PI phase. We will illustrate the FM-PI transition,
but focus more 
on the `metallic' phases, of relevance
to the disordered Re$_{1-x}$Sr$_x$MnO$_3$ family at $x \gtrsim 0.3$. 
In these  materials
phonon effects are not significant
and the effect of spin 
fluctuations on diffusive charge dynamics 
can be directly probed.

We study the following model: 
\begin{equation}
H = \sum_{\langle ij \rangle} t_{ij} 
c^{\dagger}_{i \sigma} c^{~}_{j \sigma}
+  \sum_{i } (\epsilon_i - \mu) n_{i}  
- J_H\sum_i {\bf S}_i {\bf .} {\vec \sigma}_i  
\end{equation}
The $t_{ij}=-t$ are
nearest neighbour hopping on a cubic lattice and the on site 
disorder $\epsilon_i$ is distributed uniformly between 
$\pm \Delta/2$.  
We set 
$J_H/t \rightarrow \infty$.
The parameters in the problem are  $\Delta/t$,
and density $n$ (or chemical potential $\mu$). 
We set $t=1$, fixing our basic energy scale, and assume the core
spins to be classical, with  
$\vert {\bf S}_i \vert =1$. 

There are two key steps in solving for the magnetic and transport 
properties of a  model like this: $(i)$~Evaluate 
the correlated spin distribution,
$P\{{\bf S }_i \}$, 
controlling  the magnetic response, 
by `integrating out' the
electronic degrees of freedom. $\{ {\bf S}_i \}$ 
denotes the  
full spin configuration.
$(ii)$~Solve for charge dynamics;
resistivity, optical response, etc, in the background of
structural disorder and equilibrium spin configurations.

Previous studies of the model have used 
variational mean field (VMF) theory \cite{varma}
and DMFT \cite{freericks,auslend} to access the magnetism.
Transport properties have been analysed within DMFT \cite{freericks}
and by using 
`scaling theory' \cite{sheng}
in the limits of a spin polarised $(T=0)$ state and
a fully spin disordered 
$(T \gg T_c)$ state. Finally, real

% -------------------------------------------------------------------
\begin{center}
\begin{figure}

\epsfxsize=7cm \epsfysize=5.0cm \epsfbox{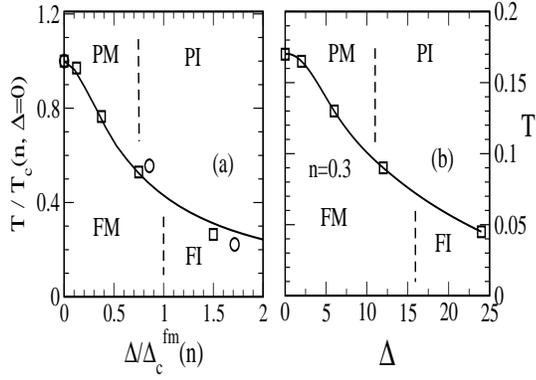}

\vspace{2mm}

\caption{$(a)$.~Approximate `universal' phase diagram of the
DDE model. 
For a specified electron density, $n$, $\Delta$  
is measured in units of the $T=0$ critical disorder,
$\Delta_c^{FM}(n)$,  
while $T$ is measured in units of $T_c(n, \Delta=0)$. 
Our data for $n=0.1$ (circles) and $n=0.3$ (squares)
are described approximately by a common fit. $(b)$.~Actual 
phase diagram for $n=0.3$ The `boundary' 
separating the PM and PI phases is a crossover.
}
\vspace{2mm}
\end{figure}
\end{center}

%--------------------------------------------------------------------

\noindent
space MC  
technique
has been used \cite{dex-dis-mc}
  to study the magnetic transition and the spin wave
spectrum.
Both VMF and DMFT `factor' the correlated spin distribution
into a product of single site distributions. 
This loses out on spin correlations and 
trivialises the paramagnetic phase.
In addition,  
DMFT misses out key vertex corrections in
 the `disorder average' process, losing 
electron  localisation effects.  
The `scaling 
theory' approach  to electronic
transport is exact for  
$T \rightarrow 0$  
and $T \gg T_c$, but not of much use in the crucial regime around $T_c$.
The only approach which implements step $(i)$ exactly 
is Monte Carlo, but the computational cost of these simulations
limits available sizes to $\sim 4^3$ in three dimension \cite{mc-large}. 
The small linear dimensions, and the large finite
size gaps, make it impossible to reliably estimate 
transport properties. This is where our method, below,  allows
a breakthrough.

For  $J_H/t \rightarrow \infty$, a 
standard transformation and projection \cite{sk-pm-mc}
can be used to map on $H$ to a `spinless fermion' problem 
with core spin dependent hopping amplitudes:
$ H \equiv   
 -t\sum_{\langle ij \rangle} f_{ij}
(~e^{i \Phi_{ij}}  \gamma^{\dagger}_i  \gamma_j + 
h.c~) + 
\sum_i (\epsilon_i - \mu) n_i $.
The hopping amplitude $g_{ij} = f_{ij} e^{i\Phi_{ij}}$, 
between locally aligned states,
can be written in terms of the polar angle $(\theta_i)$ and
azimuthal angle $(\phi_i)$ of the spin ${\bf S}_i$ 
as,
$  cos{\theta_i \over 2} cos{\theta_j \over 2}$ 
$+
sin{\theta_i \over 2} sin{\theta_j \over 2}
e^{-i~(\phi_i - \phi_j)}$.
The `magnitude'  of the overlap,
$f_{ij} = \sqrt{( 1 + {\bf S}_i.{\bf S}_j)/2 }$,
and  the phase is specified by 
$tan{\Phi_{ij}} = Im(g_{ij})/Re(g_{ij})$.

The fermions in this `quadratic' problem move 
in the background of quenched disorder $\epsilon_i$ and
`annealed disorder' in the hopping amplitudes $g_{ij}$. 
To exploit the `non interacting' character of the fermion problem we
need to know the relevant
 $\{ f , \Phi\}$ configurations, controlled by
$ H_{eff} \{ f, \Phi \} = 
 -{1 \over \beta} logTr e^{-{\beta} H}$. 
The corresponding Boltzmann distribution  
is  
$
P \{ f, \Phi  \} \propto e^{- H_{eff}\{ f, \Phi \}}$.
Our key proposal \cite{sk-pm-mc} is 
$
 -{1 \over \beta} logTr e^{-{\beta} H} \approx
-\sum_{\langle ij \rangle} D_{ij} f_{ij}  
$
The effective `exchange' $D_{ij}$, in the
short range classical spin model, 
is determined self consistently as 
the thermal average of the `mixed' spin-fermion operator,  
${\hat \Gamma}_{ij}
= (e^{i \Phi_{ij}}  \gamma^{\dagger}_i  \gamma_j + h.c)$
over the assumed equilibrium distribution.
The self-consistent $D_{ij}$ are solved for via MC for
a specified
$\mu$, $\{ \epsilon_i \}$, and $T$.
The   $D_{ij}$ can be spatially strongly inhomogeneous
and also
significantly temperature dependent.
At self-consistency fermionic averages are computed over equilibrium spin
configurations.
We work at constant $n$, fixing $\mu$ through iteration.
Since we avoid the expensive spin update procedure of `exact' MC and use
diagonalisation only to compute the $D_{ij}$ we can access
sizes $\sim 10^3$, compared to $4^3$
in the standard approach.
Transport properties, at equilibrium, 
are computed exactly 
using the Kubo formula, 
employing sizes $\sim 8^3 - 10^3$.
We  systematically 
check for size dependence in our transport results.
The conductivity results are in units of $(\pi e^2)/{\hbar}a_0$,
the Mott limit being  $(0.03 e^2)/{\hbar}a_0$.

Fig.1 shows the  
``global'' 
phase diagram of the DDE model. 
 We have studied the problem at 
$n =  0.3$ and $n =  0.1$, varying disorder from the perturbative end 
to the localisation
regime.
Fig.1.$(a)$. superposes the results at $n=0.3$ and $n=0.1$, 
appropriately scaling 
the disorder and temperature (see caption).
The critical disorder for the PM-PI crossover, or the disorder
dependence of $T_c$ need not be `universal' 
but seems to follow the same overall trend 
at moderate  $n$.
Fig.1.$(b)$ shows the `true' phase diagram specifically 
at $n=0.3$. 
The $T_c$ of DE models 
is approximately related to the internal energy change:
$T_c log(2S + 1)  \sim {\cal E}(T_c) - {\cal E}(0)$, which in turn
is related to the kinetic energy, $K$, at $T=0$.
At small $\Delta$ and $T=0$, $K(\Delta)  \sim K(0)
-  \chi_d \Delta^2$, where $\chi_d$ is the `local' 
density response function.
At strong disorder, in the
`localised' phase, $K(\Delta) \propto t^2/\Delta$. The correspondence
of these limits with the inferred $T_c(\Delta)$ is visible in 
Fig.1.$(b)$.

% -------------------------------------------------------------
\begin{center}

\vspace{.3cm}

\begin{figure}

\epsfxsize=7cm \epsfysize=5.5cm \epsfbox{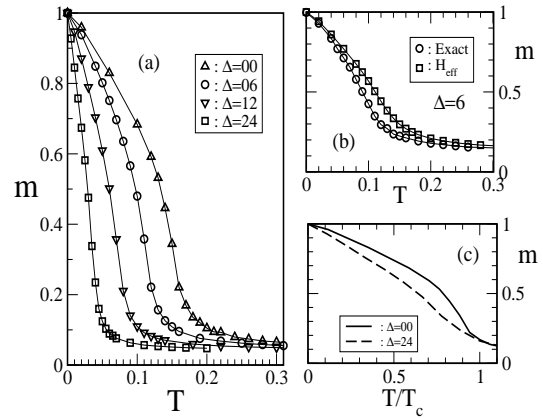}

\vspace{2mm}

\caption{$(a).$~Temperature dependence of 
magnetisation at $n=0.3$, with $\Delta$ increasing from $0-24$.
The inferred  $T_c(\Delta)$
is shown in Fig.1. $(b).$~Comparison of $m(T)$ obtained via exact MC and
$H_{eff}$ on a $4 \times 4 \times 4$ system at $n=0.3$. 
Disorder average over 8 realisations in both case.
$(c).$~The scaled magnetisation $m(T/T_c(\Delta))$, at $n=0.3, 
$for $\Delta =0$ and
$\Delta=24$.}
\vspace{2mm}
\end{figure}
\end{center}

\begin{center}

\begin{figure}

\epsfxsize=7.0cm \epsfysize=5.50cm \epsfbox{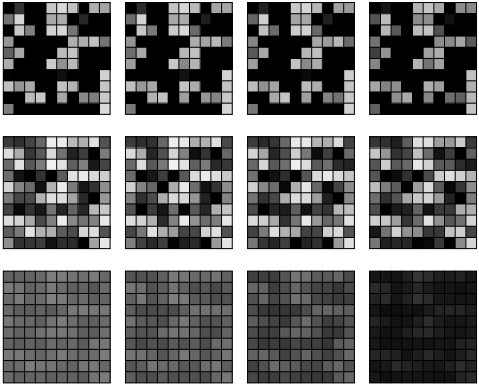}

\vspace{2mm}

\caption{Real space images, top surface of a $10 \times 10 \times 10$
cube with PBC, at $n=0.3$ and $\Delta = 6$, for a specific realisation of
disorder. First row:  
electron density distribution, 
$\langle n({\bf r}) \rangle_T$,
second row:
effective exchange, $D({\bf r},
{\bf r} + {\bf \delta}: T)$, third row:
nearest neighbour
spin correlation, 
$f_2= \langle {\bf S}({\bf r}).{\bf S}({\bf r} +{\bf \delta})
\rangle_T$.
Temperatures along the row are,
$T/T_c \sim 1.2, 0.9, 0.6 $ and $0.2$.
Dark regions correspond to high density (top row), strong exchange
(central row), and strong FM correlation (bottom).}

\end{figure}

\end{center}

% -------------------------------------------------------------

Note that even at strong localisation, $\Delta \sim 24$, local excursion 
of the electrons can still sustain a $T_c \sim 0.05$, which would
be $\sim 75 $K for $t \sim 150$meV.
DMFT and VMF also provide 
qualitatively similar
trends in $T_c$.

Fig.2.$(a)$ shows  $m(T)$ with increasing disorder, at $n=0.3$,
from which the data in Fig.1.$(b)$  was inferred, while
Fig.2.$(b)$ compares the $m(T)$ computed with $H_{eff}$
with the `exact' MC  at $\Delta=6$. 
Although the asymptotic behaviour of $T_c(\Delta)$ is easy to 
motivate,
Fig.2.$(c)$ provides the first indication that the 
properties of the  disordered ferromagnet
cannot be understood by merely scaling $T_c$. 
Even though the $T=0$ 
state is fully polarised, at low finite $T$ 
the `weak' bonds lead quickly to local disordering. The  
magnetisation $m(T/T_c)$  falls  more 
sharply at low temperature in the disordered system \cite{amorph-fm}.
The inhomogeneous character, and  
correspondingly wide distribution, of the `exchange' 
leads to an overall increase in the density of low energy 
magnetic excitations. In addition to suppressing
$m(T)$ it enhances the  
specific heat,
$C_V(T/T_c)$, and reduces the spin wave stiffness $D_{SW}$.

The real space images, Fig.3, illustrate the inhomogeneous freezing 
into a ferromagnetic state in a `cooling sequence', left to right.
The density profile $\langle n({\bf r})\rangle $, first row, 
remains  
unchanged over the $T$ range of interest, $0-1.2 T_c$.
The exchange, $D_{ij}(T)$, central row,
is strongly spatially inhomogeneous, but 
as a whole not strongly $T$ dependent. The bond
distribution, $P(D, T)$, 
reveals \cite{sk-pm-dis2} 
 that weak bonds at low $T$ get quickly weakened 
with increasing $T$ while strong 
bonds are essentially $T$ independent.
The nearest neighbour spin correlation, $f_2$, highlights 
the inhomogeneous disordering of the polarised state
with increasing $T$ (white regions, two panels on the right),
and the surviving local order for $T > T_c$ (left panel).

These  results suggest the possible  
correlation between spatial inhomogeneity in 
the magnetic state and the bulk thermodynamics. The interplay of
thermal spin disorder and `frozen' structural disorder also 
affects the charge dynamics.
Increasing  $\Delta$ increases the 
residual resistivity,  
with $\rho_0 \sim \rho_{Mott} $
at $\Delta \sim 10$. 
The resistivity in the paramagnetic phase $(\rho_{inf})$, 
tracked at $T=0.4$ in Fig.4.$(b)$, 
is not simply the additive contribution of structural
and spin disorder. If Mathiessens rule were obeyed, $\rho_{inf} -
\rho_0$  
should have been constant. The deviation arises from 
interference between structural and magnetic scattering
and is clearly observed in the metallic
Re$_{1-x}$Sr$_x$MnO$_3$, at $x=0.4$, with Re being, La, Pr, Nd
\cite{saitoh},
and in 
La$_{1-x}$Sr$_x$Mn$_{1-z}$Al$_z$O$_3$, at $x=0.3$, varying $z$
\cite{sawaki}.
This `interference'  is
beyond Boltzmann  theory. 
With $a_0 \sim 4 \AA$, 
$\rho_{Mott}  \sim
5~m\Omega$cm in the manganites. For the `disordered' LaSr family, 
as $\rho_0/\rho_{Mott}$
varies from $0.005 - 0.04$ with increasing disorder,  
the corresponding $\rho_{inf}/\rho_{Mott} $ increases from
$0.4 - 1.5$ \cite{sawaki,saitoh}.

Fig.4.$(a)$ shows the
 normalised resistivity $\delta \rho(T) =
(\rho(T) - \rho(0))/
(\rho(0.4) - \rho(0))$ for $\Delta=0,2,6$.
The shift in $T_c$ 
in this disorder regime is quite small, but 
$\delta\rho(T/T_c)$ quickly changes character with
increasing disorder.
The more prominent  short range spin fluctuations  
in the disordered system couple to the diffusive electrons 
leading  to a sharper rise in $\rho(T)$. 
While most of the rise in the `clean' system occurs in the
vicinity of $T_c$, the rise is spread over a wide interval
in the disordered system.

% ---------------------------------------------------------------------
\begin{center}
\begin{figure}
\vspace{2mm}
\epsfxsize=8.5cm \epsfysize=6.2cm \epsfbox{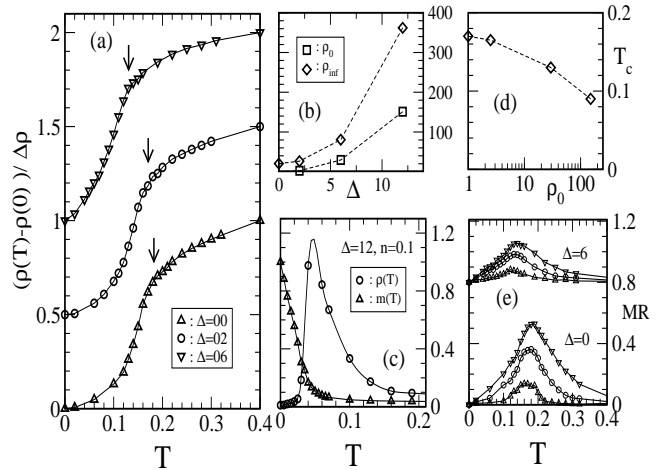}
\vspace{2mm}
\caption{ $(a).$~Temperature dependence of resistivity 
at $n=0.3$. We plot $ \rho(T, \Delta) - \rho(0, \Delta)$ 
normalised by $\rho(0.4, \Delta)
- \rho(0, \Delta)$, shifting successive curves by 0.5 for clarity.
$T_c$ marked by an arrow on each curve.
$(b).$~Resistivity at $T=0$, $\rho_0$, and at $T=0.4$, 
$\rho_{inf}$, with increasing disorder.
Note the 
clear violation of Mathiessens rule even at $\Delta =2$ (see text).
$(c)$~Resistivity normalised to maximum value $(\sim 10^4)$, 
and $m(T)$ 
at $n=0.1$ and $\Delta = 12$ 
illustrating  the FM $\rightarrow $ PI transition. 
$(d).$~Variation in
$T_c$ with residual resistivity. Note the logarithmic
$x$ scale.
$(e).$~The magnetoresistance $(\rho(T,0) - \rho(T,h))/\rho(T,0)$
at $n=0.3$: $\Delta =0$ (lower set) and $\Delta =6$ (upper set).
The $\Delta =6$ set has been vertically shifted by $0.8$. 
Field values
are $h=0, 0.02, 0.05, 0.10$.
}
\vspace{2mm}
\end{figure}
\end{center}

\begin{center}
\begin{figure}
\vspace{2mm}
\epsfxsize=8.5cm \epsfysize=6.0cm \epsfbox{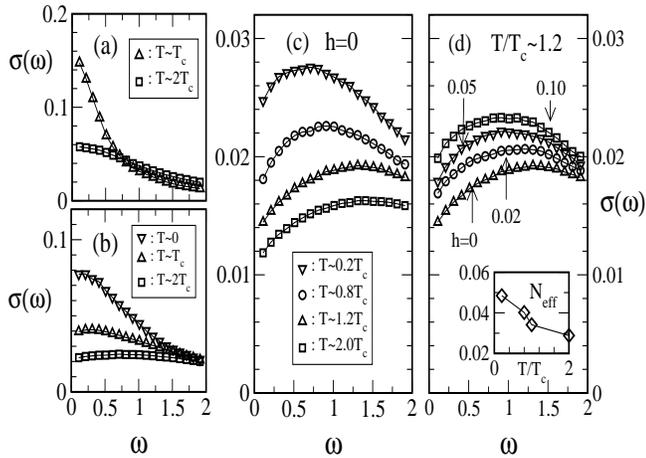}
\vspace{2mm}
\caption{Optical conductivity at $n=0.3$. $(a)$. $\Delta =0$,
$(b)$. $\Delta=4$, $(c)$. $\Delta=6$, at $h=0$, and $(d)$.
$\Delta=6$ at $T \sim T_c$. Other parameters indicated in the figure.
Inset in panel $(d)$ shows the
integrated spectral weight, $N_{eff}(T)$,
 (see text) derived from the data in panel
$(c)$.
Data averaged over $4-8$ realisations of disorder.
}
\vspace{2mm}
\end{figure}
\end{center}

% ---------------------------------------------------------------------

Fig.4.$(c)$. shows data at $n=0.1$ and $\Delta=12$ to illustrate  the
FM-PI transition.  We work with this lower density
because the MIT is easier to access.  We solve the full  problem  
for sizes $6 \times 6 \times L$, with $L =\{ 6,12, 24 \}$,
disorder average, and extrapolate the computed `d.c conductivity' to
$L \rightarrow \infty$. 
This is crucial to capture `Anderson localisation' in finite systems.
The transport
in the insulating phase is controlled by activation to the
mobility edge.

The correlation between $T_c$ and residual resistivity has been 
experimentally explored \cite{coey,attfield}
and Fig.4.$(e)$ highlights the fall in $T_c$
as $\rho_0$ heads towards the Mott limit. 

Our results on MR, Fig.4.$(e)$,
indicate that in the  intermediate disorder regime 
the MR is actually {\it smaller}
in the disordered problem than in the clean system. 
The MR is 
$ 1  - \rho(T,h)/\rho(T,0)$.
The minimum $\rho(T,h)$ is limited by $\rho_0$, the residual resistivity.
With increasing $\Delta$, $\rho_0$ grows, limiting the MR. The
MR rises again only when 
$\Delta$ is large enough to drive a finite $T$, FM-PI
transition.

Finally, the optical response of the system,  at $n=0.3$ and $\Delta =0, 4$ and
$6$, is shown in Fig.5. 
Panel $(a).$ shows $\sigma(\omega)$ at  $\Delta=0$
and the Drude feature survives even for $T \sim 2 T_c$ with 
$\sigma_{dc} \sim  6 \sigma_{Mott}$. 
For $\Delta=4$, however, panel $(b)$. reveals that $\sigma(\omega)$ 
changes from a Drude form for $T \lesssim T_c$ 
to an essentially `flat' incoherent 
response at $T \sim 2T_c$. At $2T_c$, $\sigma_{dc} \sim 2 \sigma_{Mott}$.
This response is roughly like La$_{1-x}$Sr$_x$MnO$_3$ at $x \sim 0.4$
\cite{saitoh}.
At even larger disorder, $\Delta=6$, the response is non Drude 
{\it even at $T=0$}, and becomes markedly so, with a finite $\omega$
peak, as $T$ is increased. 
The apparently $\sqrt \omega$ rise  results from the intimate
coupling of the diffusive electrons to spin fluctuations via the hopping
modulation.  
For $\Delta=6$, $\sigma_{dc} \sim 2\sigma_{Mott}$ at $T=0$, and
$\sigma_{dc} \sim \sigma_{Mott}$ at $T \sim 2T_c$.
This is like the response in 
(Pr,Nd)$_{1-x}$Sr$_x$MnO$_3$ at $x \sim 0.4$ \cite{saitoh}. 
We predict that the 
response will be similar at $x=0.4$ in  
La$_{1-x}$Sr$_x$Mn$_{1-z}$Al$_z$O$_3$ at $z \sim 1\% - 2\%$. 
The inset to panel $(d)$. shows the integrated low 
frequency spectral weight 
$N_{eff}(\omega', T) = \int_0^{\omega'}
\sigma(\omega, T) d \omega$, at $\omega' = 2$. There is a $40 \%$ loss
in spectral weight in the $T$ range $0-2T_c$.
Our energy cutoff in $N_{eff}$ 
roughly corresponds 
to $\omega' \sim 0.3$eV in the manganites.
Panel $(d).$ shows the magneto-optical response at $\Delta=6$ and
$T \sim 1.2 T_c$.

The non Drude relaxation with $\sigma(\omega)$ having a finite $\omega$
peak occurs in a regime where $d\rho/dT > 0$, as in conventional metals,
but the charge dynamics is highly diffusive as expected in a system
with strong `effective disorder'. The change from `coherent'
to `incoherent' dynamics occurs in $\sigma(\omega, T)$ when $\rho_{dc}(T)
\sim \rho_{Mott}$. This general feature is true of the PM 
phase of all the manganites.

In conclusion, we have discussed the 
inhomogeneous magnetism in disordered double exchange magnets,
and provided the first ``exact'' results on transport and
optical response. 
The dependence of the  non Drude relaxation 
on disorder and temperature 
is fully consistent with the metallic manganites.
The extension of our method to include phonons 
will allow a complete solution of charge response in 
the  manganites.

We acknowledge use of the Beowulf cluster at H.R.I.

{}

% -------------------------------------------------------------------

\end{document}